\documentclass{PoS}

\newcommand{\be}{\begin{equation}}
\newcommand{\ee}{\end{equation}}
\newcommand{\bea}{\begin{eqnarray}}
\newcommand{\eea}{\end{eqnarray}}
\newcommand{\non}{\nonumber}

\newcommand{\mbf}{\mathbf}

\title{The chiral crossover, static-light and light-light meson spectra, and the deconfinement crossover}

\ShortTitle{quark mass and meson spectrum at deconfinement}

\author{\speaker{Pedro Bicudo}\thanks{I thank the organizers for maintaining this beautiful meeting in the spirit of Ileana Iori.}\\
        CFTP, Dep. F\'{\i}sica, Instituto Superior T\'ecnico,
Av. Rovisco Pais, 1049-001 Lisboa, Portugal\\
        E-mail: \email{bicudo@ist.utl.pt}}

        \author{Nuno Cardoso \\
        CFTP, Dep. F\'{\i}sica, Instituto Superior T\'ecnico,
Av. Rovisco Pais, 1049-001 Lisboa, Portugal\\
        E-mail: \email{nunocardoso@cftp.ist.utl.pt}}

        \author{Marco Cardoso \\
        CFTP, Dep. F\'{\i}sica, Instituto Superior T\'ecnico,
Av. Rovisco Pais, 1049-001 Lisboa, Portugal\\
        E-mail: \email{mjdcc@cftp.ist.utl.pt}}
        
\abstract{We study the chiral crossover, the spectra of light-light and of static-light mesons and the deconfinement crossover at finite temperature T.  Our framework is the confining and chiral invariant quark model, related to truncated Coulomb gauge QCD. Since we are dealing with light quarks, where the linear potential dominates the quark condensate and the spectrum, we only specialize in the linear confining potential for the quark-antiquark interaction. We utilize T dependent string tensions previously fitted from lattice QCD data, and a fit of previously computed dynamically generated constituent quark masses. We scan the T effects on the constituent quark mass, on the meson spectra and on the polyakov loop.}

\FullConference{XLIX International Winter Meeting on Nuclear Physics, BORMIO2011\\
		January 24-28, 2011\\
		Bormio, Italy}

\begin{document}

\section{Introduction}

\begin{figure}[t]
\hspace{.5cm}
\center{
\includegraphics[angle=180,width=0.9\columnwidth]{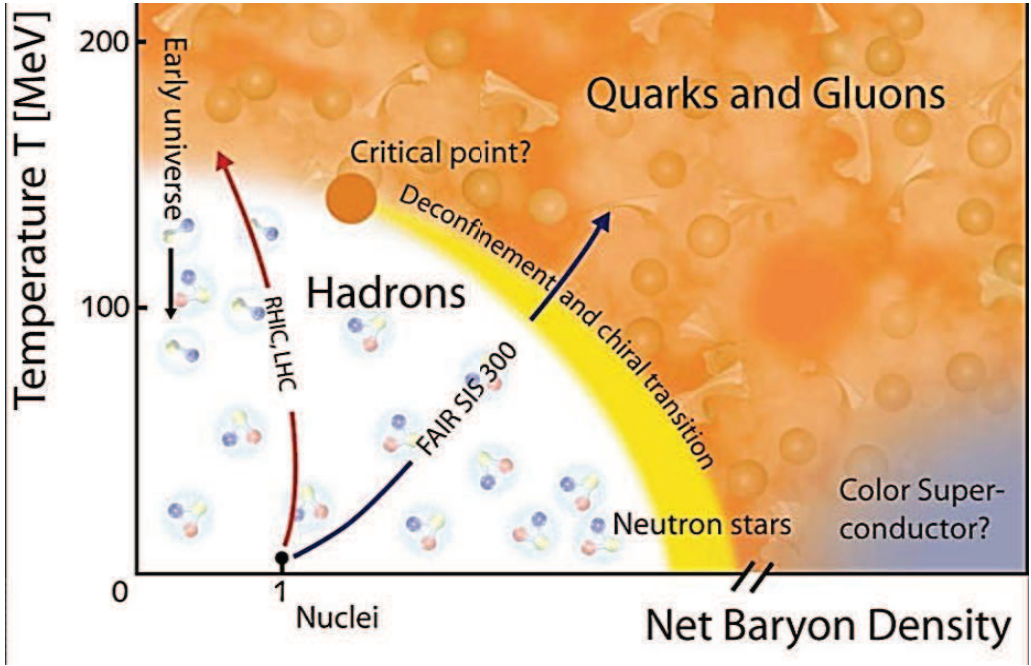}
}
\caption{A sketch of the QCD Phase Diagram, according to
the collaboration CBM at FAIR
\cite{CBM}, it is usually assumed that the critical point for deconfinement
coincides with the critical point for chiral symmetry restoration.}
\label{CBM}
\end{figure}

Our main motivation is to contribute to understand the QCD phase
diagram 
\cite{CBM}, 
for finite $T$ and $\mu$. The QCD phase diagram is scheduled to be studied at LHC, RHIC and FAIR,
and is sketched in Fig \ref{CBM}.
Notice that, after enormous theoretical efforts, the analytic crossover nature of the finite-temperature QCD transition 
was finally determined by Y. Aoki {\em et al.}
\cite{Aoki:2006we},  
utilizing Lattice QCD and physical quark, and reaching the continuum 
extrapolation with a finite volume analysis.

\begin{figure}[t]
\vspace{1cm}
\center{
\includegraphics[width=0.9\columnwidth]{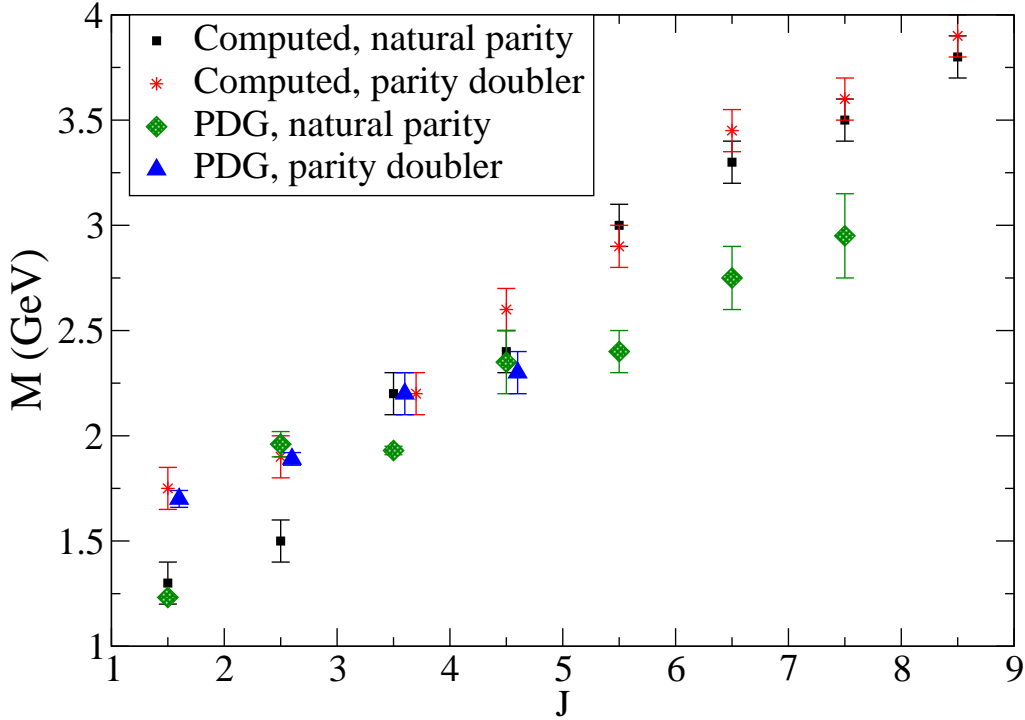} 
}
\caption{ First calculation of excited baryons with a chiral invariant quark model
\cite{Bicudo:2009cr}.}
\label{excitedBaryons}
\end{figure}

Here we utilize the Coulomb gauge hamiltonian formalism of QCD,  presently the
only continuum model of QCD able to microscopically include both a quark-antiquark confining potential and
a vacuum condensate of quark-antiquark pairs. This model is able to address excited
hadrons as in Fig. \ref{excitedBaryons},
and chiral symmetry at the same token, and we recently suggested that
the infrared enhancement of the quark mass can be observed in the excited
baryon spectrum at CBELSA and at JLAB
\cite{Bicudo:2009cr,Bicudo:2009hm}.
Thus the present work, not only addresses the QCD phase diagram,
but it also constitutes the first step to allow us in the future to
extend the computation of nay hadron spectrum, say the Fig. \ref{excitedBaryons} computed in
reference \cite{Bicudo:2009cr}, to finite $T$ .

Both for the study of the hadron spectra,
and for the study the QCD phase diagram,
a finite quark mass is relevant.
In the phase diagram, a finite current quark mass $m_0$
affects the position of the critical point between the 
crossover at low chemical potential $\mu$ and the phase transition
at higher $\mu$. Moreover the current quark mass
affects the QCD vacuum energy density $\cal E$, relevant
for the dark energy of cosmology.
This all occurs in the dynamical generation of the quark 
mass $m(p)$.
While the quark condensate $ \langle \bar \psi \psi \rangle$
is a frequently used order parameter for chiral symmetry breaking,
the mass gap, {\em i. e.    } the quark mass at vanishing
momentum $m(0)$ is another possible order parameter for
chiral symmetry breaking. 

\begin{figure}[t]
\hspace{.5cm}
\center{
\includegraphics[width=0.9\columnwidth]{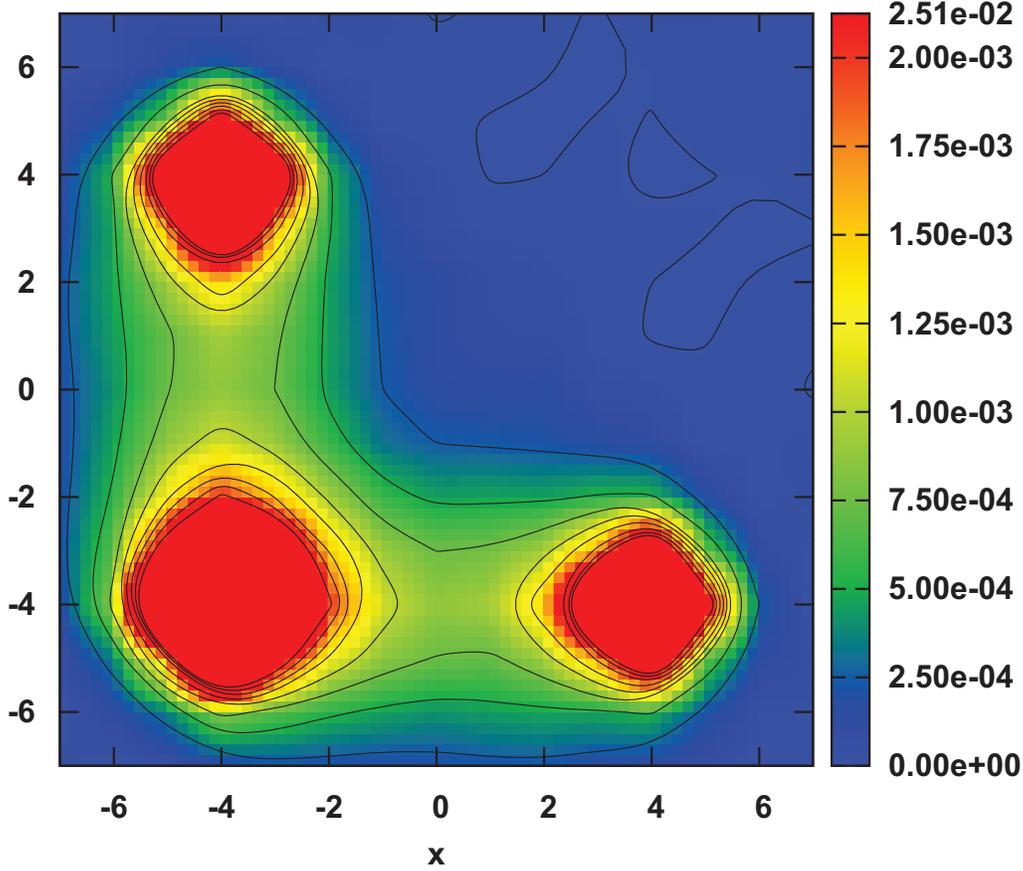}
}
\caption{The fields generated by static hybrid sources, notice how
the flux organizes into fundamental colour triplet flux tubes.}
\label{QGQfield}
\end{figure}

Here we address the finite temperature string tension, the quark mass gap
for a finite current quark mass and temperature, and the deconfinement and
chiral restoration crossovers. We conclude on the separation of the critical 
point for chiral symmetry restoration from the critical point for
deconfinement.

\begin{figure}[t!]
\vspace{1cm}
\center{
\includegraphics[width=0.8\columnwidth]{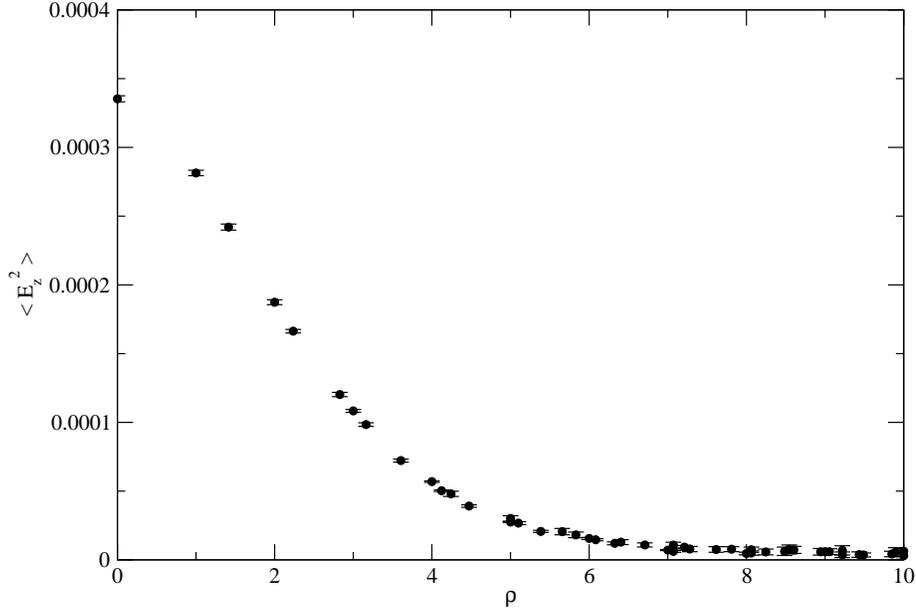}
}
\caption{The perpendicular profile of the longitudinal electric field, in units of the lattice spacing $a=0.07261(85) fm$. The flux tube is so thin, that in quark models it is usually modelled by a single parameter, the string tension $\sigma$.}
\label{Elong}
\end{figure}

\section{Fits for the finite T string tension from the Lattice QCD energy $F_1$}

At vanishing temperature $T=0$, the confinement, due to the formation of chromo-electric
and chromo-magnetic flux tubes as shown in Figs. \ref{QGQfield} and \ref{Elong},
can be modelled by a string tension, dominant at moderate distances,
\be
V(r) \simeq  {\pi \over 12 r} + V_0 + \sigma  r \ .
\ee 
At short distances we have the Luscher or Nambu-Gotto Coulomb potential due to the
string vibration plus the One Gluon Exchange Coulomb potential, 
however the Coulomb potential is not important for chiral symmetry breaking. 
At finite temperature the string tension $\sigma(T)$
should also dominate chiral symmetry breaking, and thus one of our crucial steps 
here is the fit of the string tension $\sigma(T)$ obtained
from the Lattice QCD data for the quark-antiquark free energy of the Bielefeld Lattice QCD group,
\cite{Doring:2007uh,Hubner:2007qh,Kaczmarek:2005ui,Kaczmarek:2005gi,Kaczmarek:2005zp}.

\begin{figure}[t!]
\hspace{0cm}
\center{
\includegraphics[width=0.9\columnwidth]{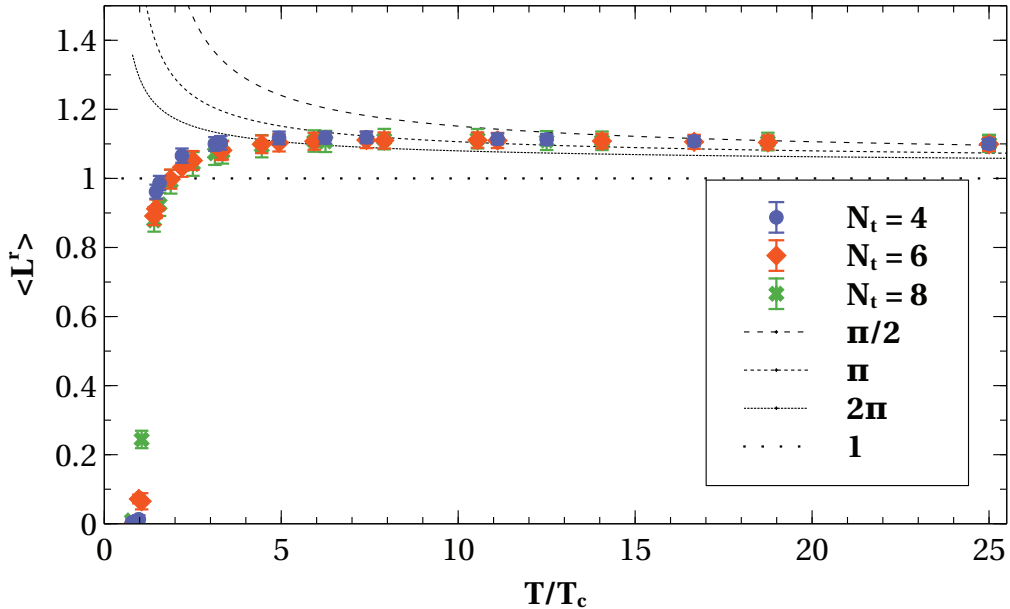}
}
\caption{
The renormalized Polyakov loop of a single quark,
computed by our Lattice QCD group.
}
\label{PtQCD}
\end{figure}

At finite temperature, the quark, or the quark-antiquark free energies, can be computed utilizing Polyakov loops, 
as in Fig. \ref{PtQCD}.
The Polyakov loop is a gluonic path, closed in the imaginary time $t_4$ (proportional
to the inverse temperature $T^{-1}$) direction  
in a periodic boundary Euclidian Lattice discretization of QCD. 
They measures the free energy $F$ of one or more static quarks,
\be
P(0) = N e^{ - F_q /T } \ , \ \ P^a(0)\bar P^{\bar a}(r) = N e^{ - F_{ q \bar q}(r)  /T }  \ .
\ee
If we consider a single solitary quark in the universe, in the confining
phase, his string will travel as far as needed to connect the quark to an antiquark,
resulting in an infinite energy F. Thus the 1 quark Polyakov loop $P$ is a
frequently used order parameter for deconfinement.
With the string tension $\sigma(T)$ 
extracted from the $q \bar q$ pair of Polyakov loops
we can also estimate the 1 quark Polyakov loop $P(0)$.

\begin{figure}[t!]
\resizebox{0.9\textwidth}{!}{%
\includegraphics{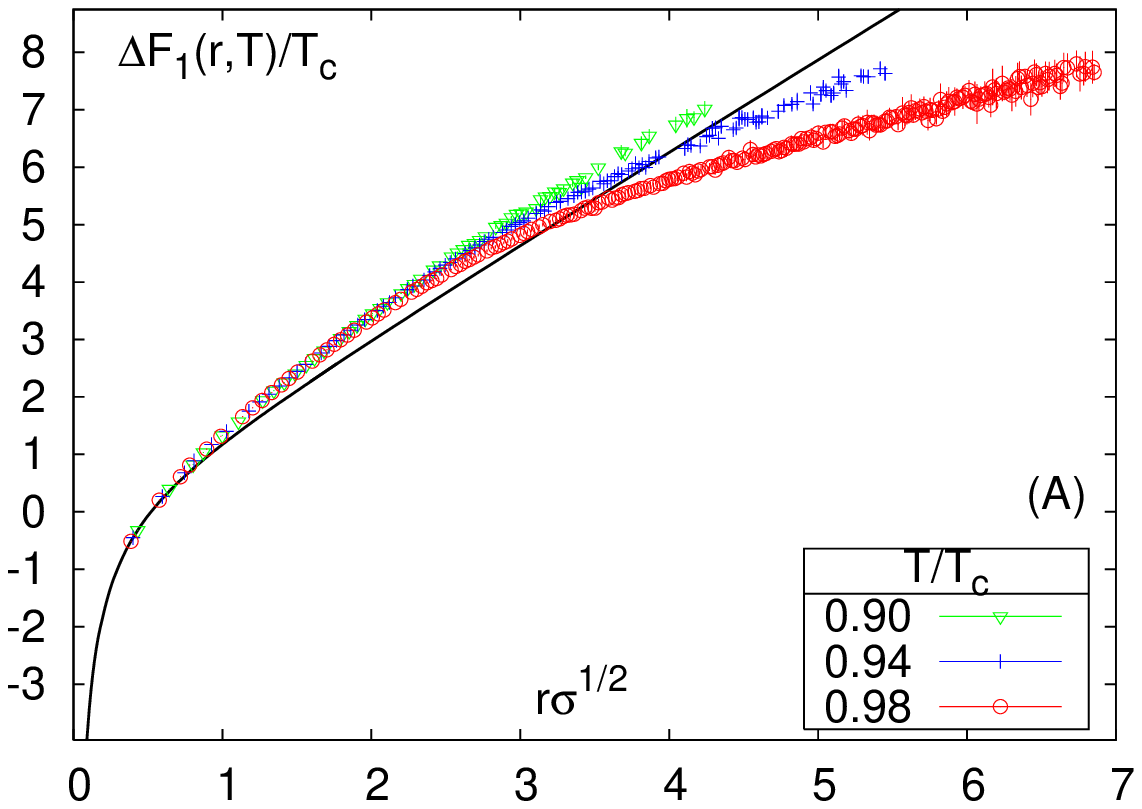}
}
\resizebox{0.9\textwidth}{!}{%
\includegraphics{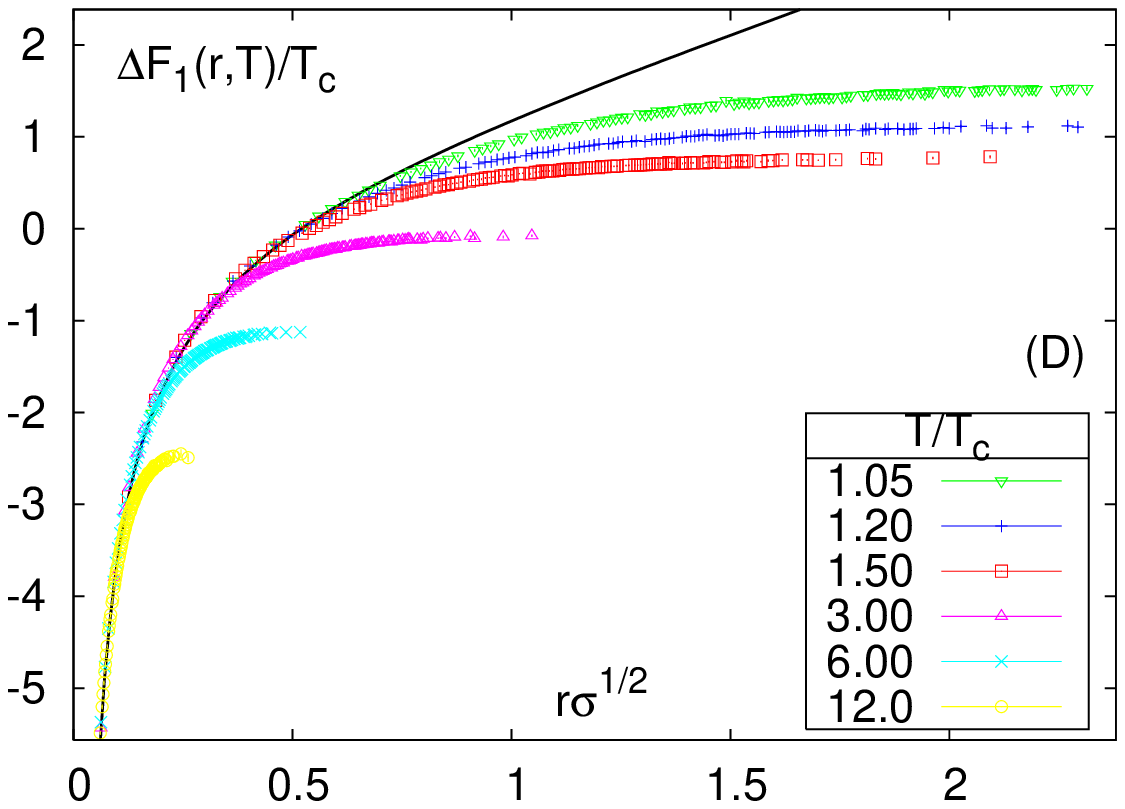}
}
\caption{\label{F1Kacz}
We show examples finite temperature static quark-antiquark potentials, in particular the  $T<T_c$  and $T>T_c$ 
Lattice QCD data for the free energy $F_1$, thanks to 
\cite{Doring:2007uh,Hubner:2007qh,Kaczmarek:2005ui,Kaczmarek:2005gi,Kaczmarek:2005zp} Olaf Kaczmarek et al.
 The solid line represents the $T=0$ static quark-antiquark potential. In this paper we discuss the use of the free energy as a finite temperature quark-antiquark potential.}
\end{figure}

\begin{figure}[t!]
\hspace{0cm}
\center{
\includegraphics[width=0.8\columnwidth]{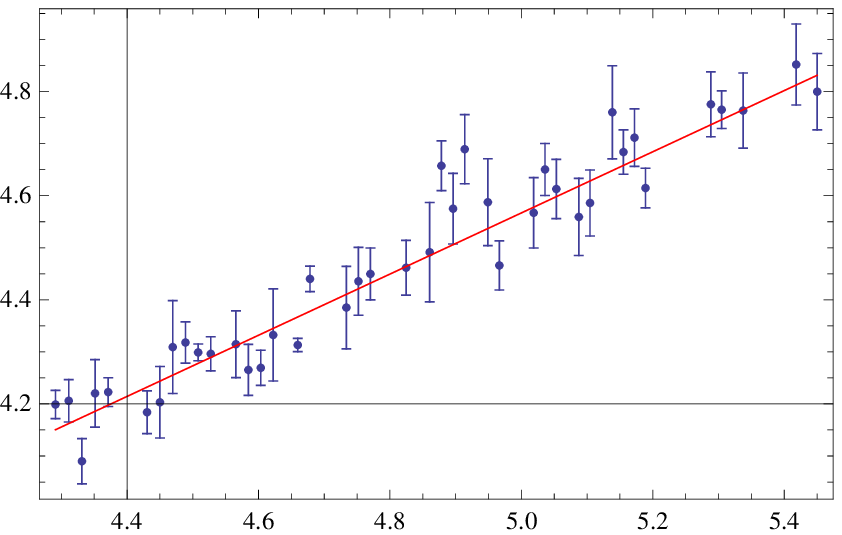}
}
\caption{
 Detail of the string tension fit in the case of $T =0.94 T_c$.
 We cut the low distance part in such a way that a linear fit is stable 
 for cuttof changes.
}
\label{fitstringtension}
\end{figure}

At finite $T$, we use as thermodynamic
potentials the free energy $F_1$ , computed in Lattice
QCD with the Polyakov loops
\cite{Doring:2007uh,Hubner:2007qh,Kaczmarek:2005ui,Kaczmarek:2005gi,Kaczmarek:2005zp},
and illustrated in Fig.  \ref{F1Kacz}.
It is related to the static potential
$V(r)   = - f d r$
with
$F1 (r)= - f d r - S d T$
adequate for isothermic transformations.
In Fig. \ref{fitstringtension} we extract the string tensions $\sigma(T)$
from the free energy $F_1(T)$ 
computed by the Bielefeld group, and we also include string tensions
previously computed by the Bielefeld group 
\cite{Kaczmarek:1999mm}.

\begin{figure}[t!]
\hspace{0cm}
\center{
\includegraphics[width=0.6\columnwidth]{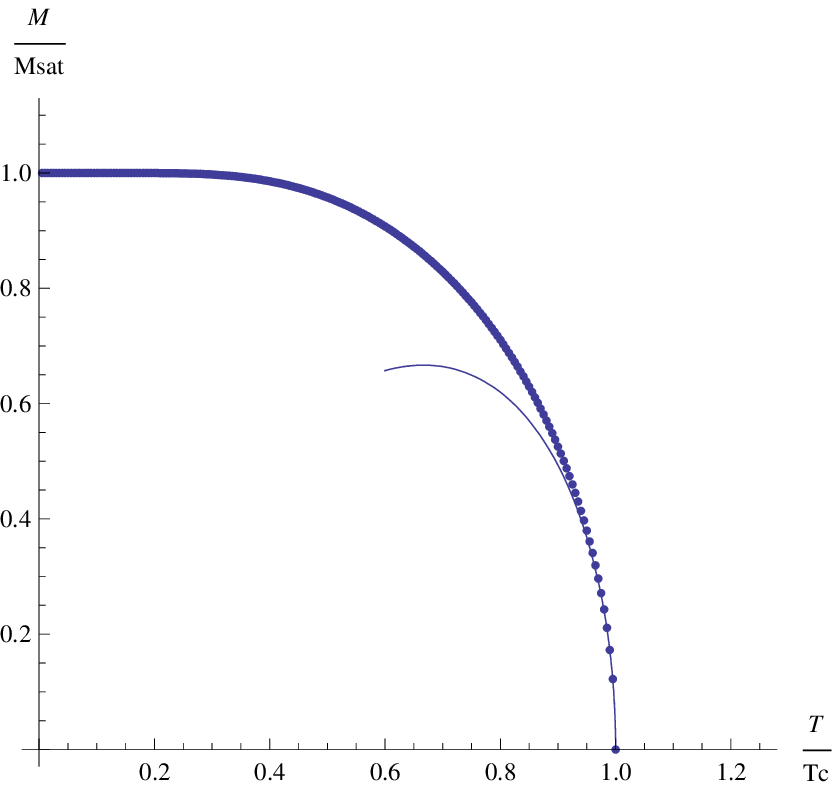}
\hspace{2cm}
\includegraphics[width=0.6\columnwidth]{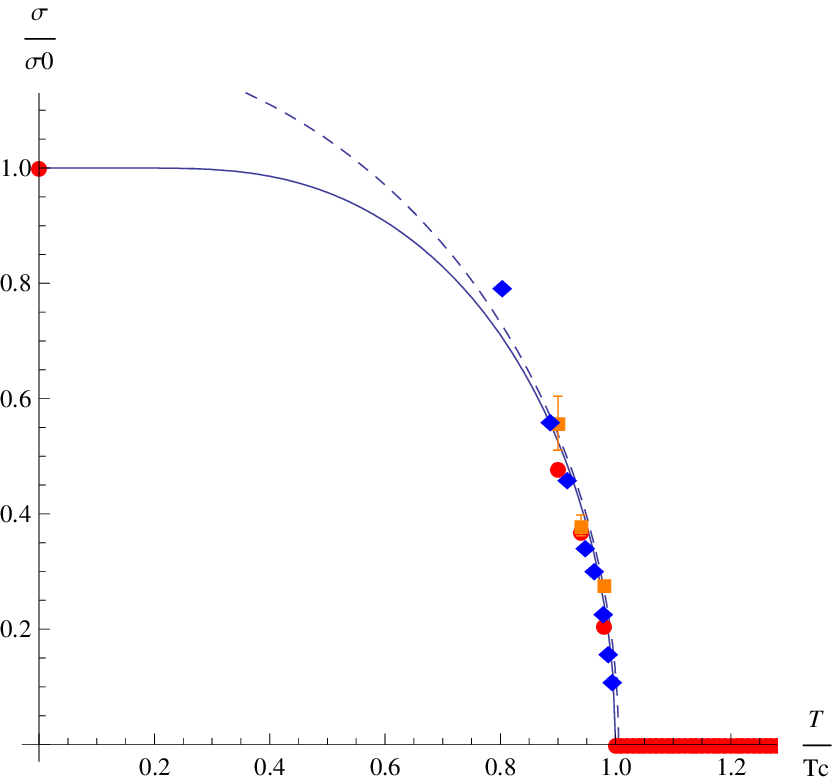}
}
\caption{
(top) The critical curve for $M \over M_{sat}$ as a function of $T \over T_c$,
for $T\simeq T_c$ it behaves like a square root. 
(bottom) Comparing the magnetization critical curve with the string tension 
$\sigma / \sigma_0$, fitted from the long distance part of $F_1$, 
they are quite close.
}
\label{magnetiz}
\end{figure}

We also find an ansatz for the string tension curve, among
the order parameter curves of other physical systems related to 
confinement, i. e. in ferromagnetic materials, in the Ising
model, in superconductors either in the BCS model or in the
Ginzburg-Landau model, or in string models, to suggest 
ansatze for the string tension curve. We find that the order parameter
curve that best fits our string tension curve is
the spontaneous magnetization of a ferromagnet 
\cite{FeynmanLS},
solution of the
algebraic equation,
\be
{M \over M_{sat}} = \tanh \left(  { T_c \over T}  {M \over M_{sat}}  \right) \ .
\label{eqformagnetization}
\ee
In Fig. \ref{magnetiz} we show the solution of Eq. \ref{eqformagnetization}
obtained with the fixed point expansion, and compare it with the
string tensions computed from lattice QCD data.

\section{The mass gap equation with finite $T$ and
finite current quark mass $m_0$.}

Now, the critical point occurs when the phase transition changes to a crossover,
and the crossover in QCD is produced by the finite current quark mass m0,
since it affects the order parameters $P$ or $\sigma$, and  the mass gap $m(0)$
 or the quark condensate $\langle \bar q q \rangle$.
 Moreover
utilizing as order parameter the mass gap, i. e. the quark mass at
vanishing moment,
a finite quark mass transforms the chiral symmetry breaking from a phase transition into a crossover.  For the
study of the QCD phase diagram it thus is relevant to determine how the current quark mass affects chiral symmetry breaking, in particular we study in detail the effect of the finite current quark mass on chiral symmetry breaking, in the framework of truncated Coulomb gauge QCD
with a linear confining quark-antiquark potential.

The most fundamental information for the quark-antiquark
interaction in QCD comes from the Wilson loop in Lattice
QCD, providing the confining quark-antiquark potential 
for a static quark-antiquark pair.
This potential is consistent with the funnel potential, also
utilized in the quark model to describe the quark-antiquark sector
of meson spectrum, in particular to describe the
linear behaviour of mesonic Regge trajectories.
Notice that the short range Coulomb potential 
could also be included in the interaction, but 
here we ignore it since it only affects the quark 
mass through ultraviolet renormalization 
\cite{Bicudo:2008kc}, 
which is assumed to be already included in the 
current quark mass. Here we specialize in computing
different aspects of chiral symmetry breaking with
linear confinement $V=\sigma r$. Since we are interested
in working at finite temperature $T$ we utilize a recent
fit of lattice QCQ data with a temperature dependent 
string tension $\sigma(T)$.

To address the light quark sector it is not sufficient to know the static 
quark-antiquark potential, we also need to know what Dirac vertex to 
use in the quark-antiquark-interaction. This vertex is necessary to
study not only the meson spectrum but also the dynamical spontaneous
breaking of chiral symmetry.
To determine what vertex to use, we review how the quark-antiquark
potential can be approximately derived from QCD, in two different gauges.
In Coulomb gauge
\cite{TDLee}, 
\be
\mathbf \nabla \cdot \mathbf A(\mbf x, t)=0 
\ee
the interaction potential, 
as derived by Szczepaniak and Swanson
\cite{Szczepaniak:1995cw,Szczepaniak:1996gb},
is a density-density  interaction, with Dirac structure
$\gamma^0 \otimes \gamma^0$.
Another approximate path from QCD 
considers the modified coordinate gauge of Balitsky 
\cite{Balitsky:1985iw}
and in the interaction potential for the
quark sector,
retains the first cumulant  order, of two gluons
\cite{Dosch:1987sk,Dosch:1988ha,Bicudo:1998bz}.
This again results in a simple density-density effective 
$\gamma^0 \otimes \gamma^0$ confining interaction. 
As in QCD, this only has one scale, say $K_0$, 
in the interaction, since both the quark
condensate and the hadron spectrum turn out to
be insensitive to any constant $U$ in the potential.
Thus our framework is similar to an expansion of
the QCD interaction, truncated to the leading 
density-density term, where the confining
quark-antiquark potential is a linear potential.

While this is not exactly equivalent to QCD,
our framework maintains three interesting
aspects of non-pertubative QCD,  
a chiral invariant quark-antiquark interaction,
the cancellation of infrared divergences
\cite{Orsay1,Orsay2,Orsay3,Orsay4,Kalinowski,Lisbon1},
and  a quark-antiquark linear potential
\cite{linear1,linear2,linear3,Szczepaniak:1995cw,linear4,Wagenbrunn:2007ie}.
Importantly, since our model is well defined and solvable, 
it can be used as a simpler model than QCD, and yet 
qualitatively correct, to address different aspects of
hadronic physics. In particular here we study how
chiral symmetry breaking occurs at
finite temperature $T$ and chemical potential $\mu$,
in the realistic case of small but finite current quark masses.
Thus we apply our framework to the phase diagram
of QCD.

Our interaction potential for the quark sector is,
\begin{eqnarray}
V_I &=& \int\, d^3x \left[ \psi^{\dag}( x) \;(m_0\beta -i{\vec{\alpha}
\cdot \vec{\nabla}} )\;\psi( x)\;+
{ 1\over 2}  \int d^4y\, \
\right.
\nonumber \\
&&
\;\psi^{\dag}( \mbf x)
\lambda^a \psi ( \mbf x)  
{-3 \over 16}  V(|\mbf x -\mbf y|)
\;\psi^{\dag}( \mbf y)
\lambda^b 
 \psi( \mbf y)  
\label{hamilt}
\end{eqnarray}
where the density-density interaction includes just the linear confining potential
together with an infrared constant, which may be possibly divergent.

\begin{figure}[t!]
\begin{center}
\includegraphics[width=0.9\columnwidth]{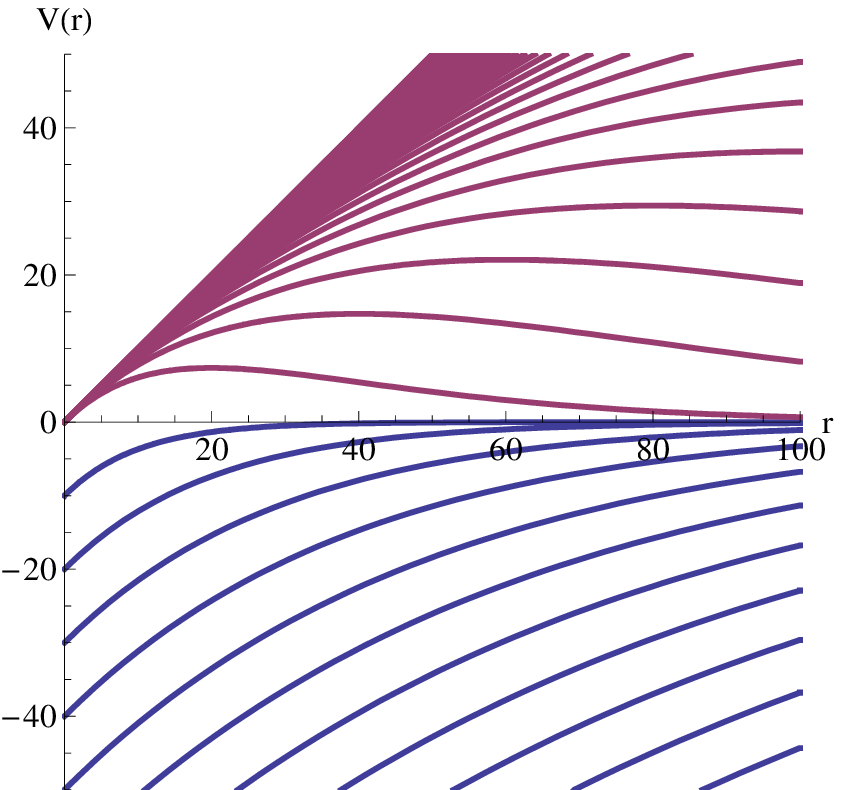}
\end{center}
\vspace{.5cm}
\caption{The linear potential must be regularized for the Fourier Transform,
and we show our two different regularizations both leading to $V(r) \to 0$ when $r \to 0$.
Two successions of curves are plotted, leading exactly to linear potentials in
the limit of a vanishing infrared regulator $\mu \to$.
The regularization of the negative curves maintains the potential monotonous,
but adds an infrared negative constant to the potential.
The regularization of the positive curves maintains $V(0)=0$ 
but the potential decreases for large $r$.
\label{linearregu}}
\end{figure}

The mass gap equation and the energy of a quark are determined
from the Schwinger-Dyson equation at one loop order using the 
Hamiltonian of Eq. (\ref{hamilt}). for a recent derivation
with all details see 
\cite{Bicudo:2010qp}.  The interaction in the four momentum
of the potential and quark propagator term includes 
an integral in the energy 
\be
\int_{- \infty}^{+ \infty} {d \, p^0 \over 2 \pi}
{ i \over p^0 - E(\mbf p) + i \epsilon}  = -{1 \over 2 } \ ,
\label{minkowski}
\ee
which factorizes trivially from the vector $\mbf p$ momentum integral.
Using spherical coordinates, the angular integrals can be performed
analytically and finally only an integral in the modulus of the momentum 
remains to be computed numerically.
We arrive at the mass gap equation in two equivalent forms,
of a non-linear integral functional equation,
\bea
\label{massgabackbacktosincos}
0 &=& p S(p) - m_0 C(p) - { \sigma \over p^2}
\int_0^\infty {d k \over 2 \pi}  \, \bigl[
\\ \nonumber 
&&
I_A(k,p,\mu)\,  S(k) C(p) 
- I_B(k,p,\mu) \,  S(p) C(k) \bigr] \ ,
\eea
and of a minimum equation of the energy density $ {\cal E}$, 
\bea
\label{energybacktosincos}
&& {\cal E} = { -g \over 2 \pi}\int_0^\infty  {dp \over 2 \pi}
 \biggl[
2p^3 C(p) + 2 p^2 m_0 S(p) + \sigma \times
\\ \nonumber 
&&
\int_0^\infty {d k \over 2 \pi} 
 I_A(k,p,\mu) \,   S(k) S(p) 
+ I_B(k,p,\mu) \,   C(p) C(k) \biggr]  \ ,
\eea
where the functions $I_B$ and $I_A$ are angular integrals
of the Fourier transform of the potential.
In what concerns the one quark energy we get,
\bea
E(p) &= &  
p C(p) + m_0 S(p)  + {\sigma \over p^2}\, \int_0^\infty {d k \over 2 \pi}  I_A(k,p,\mu) \times
\label{regularized E}
\non
\\ 
&& 
\,    S(k) S(p)  +  I_B(k,p,\mu) \,   C(p) C(k)   \ .
\eea

\begin{figure}[t!]
\begin{center}
\includegraphics[width=1.\columnwidth]{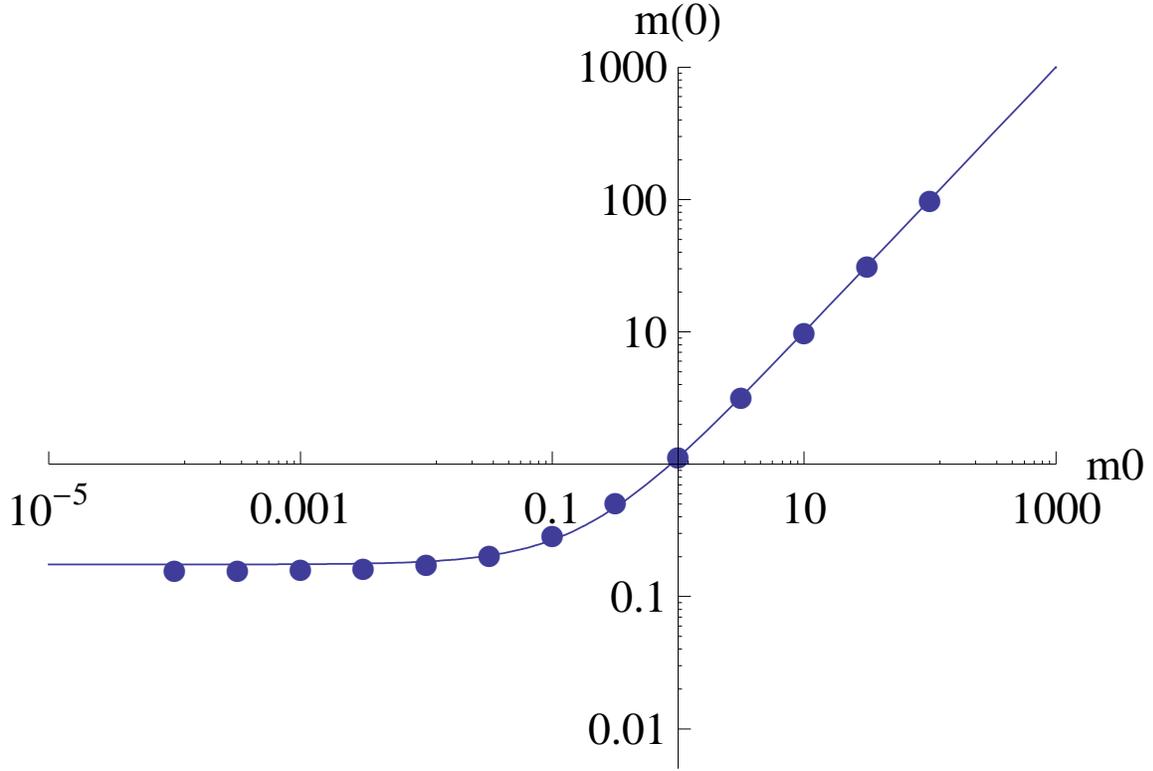}
\caption{We plot the solution of the mass gap equation $m(0)$ at $T=0$, 
for different values of the current quark mass $m_0$ in the case where the string tension
is $\sigma=1$. We also show, with a solid line, the fit with a two-parameter irrational function.
\label{fitmassgap}}
\end{center}
\end{figure}

\begin{figure}[t!]
\begin{center}
\includegraphics[width=1.\columnwidth]{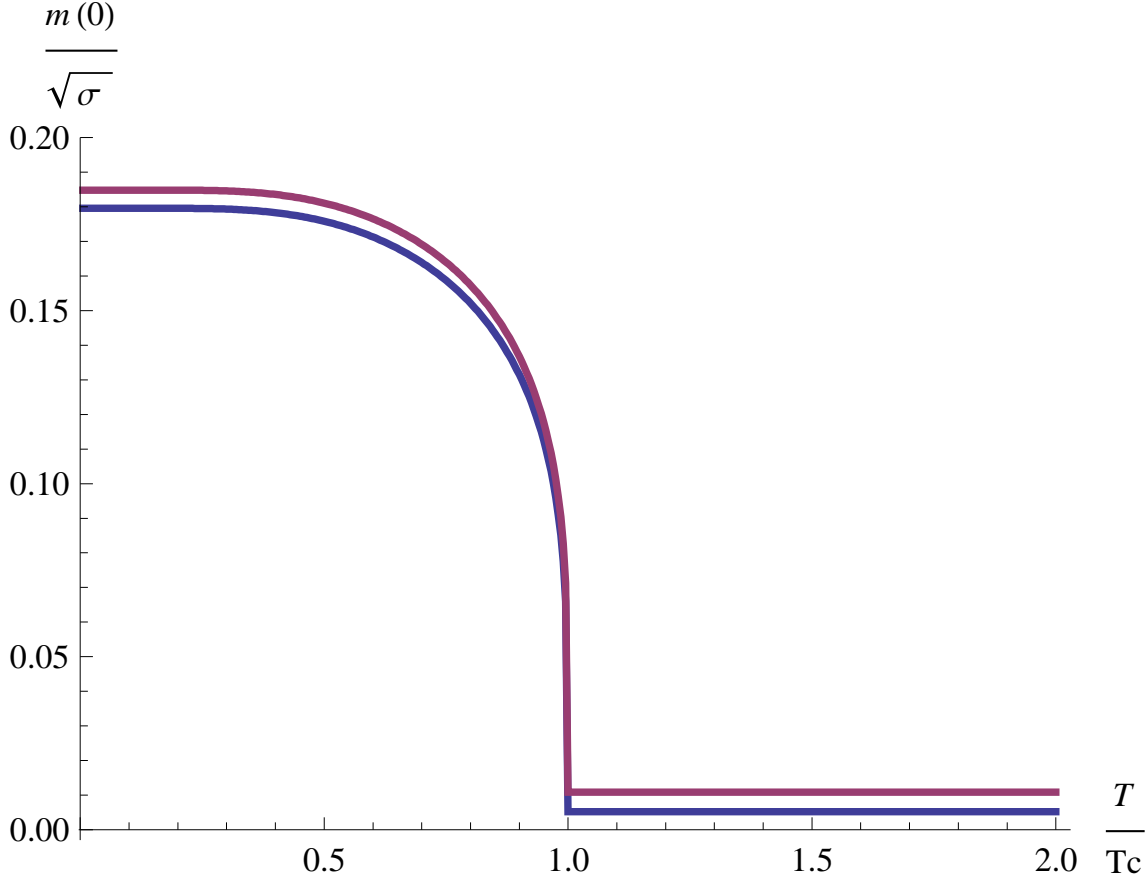}
\caption{We plot the mass gap $m(0)$ as a function of $T$ for
the light $u$ (lightest) and $d$ (slightly heavier) quarks. 
We have a crossover, close to a phase transition, since the
current masses of the light quarks are much smaller than
the dynamically generated constituent quark mass.
\label{massgapofTlights}}
\end{center}
\end{figure}

\begin{figure}[t!]
\begin{center}
\includegraphics[width=1.\columnwidth]{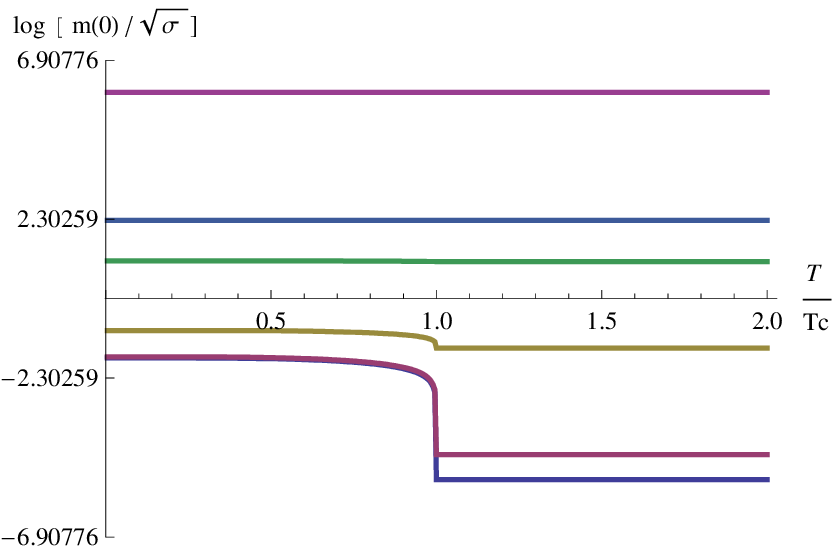}
\caption{Log plot of the mass gap $m_q(0)$ for the six
different flavours of quarks as a function of $T$. From bottom to 
top we show the quarks $u$, $d$, $s$, $c$, $b$ and $t$.
The heavier the quark, the weaker the crossover gets.
\label{massgapofT}}
\end{center}
\end{figure}

In the chiral limit of massless current quarks, the breaking of chiral symmetry is spontaneous. But for
a finite current quark mass, some dynamical symmetry breaking continues to add to the explicit breaking caused by the quark mass.  
The mass gap equation at the ladder/rainbow truncation of Coulomb
Gauge QCD in equal time reads,
\bea
\label{fixedpointeq}
&&m(p) = m_0 + { \sigma \over p^3}
\int_0^\infty {d k \over 2 \pi}  {   
 I_A(p,k,\mu) \, m (k) p   - I_B(p,k,\mu) \, m(p) k  
\over \sqrt{k^2 + m(k)^2}} \ , 
\\ \nonumber 
&&
\ I_A(p,k,\mu)=\left[
{ p k \over (p-k)^2 + \mu^2} 
- { p k \over (p+k)^2 + \mu^2} \right] \ ,
\\ \nonumber 
&&  
\ I_B(p,k,\mu)= \left[
{ p k \over (p-k)^2 + \mu^2} 
+ { p k \over (p+k)^2 + \mu^2} + {1 \over 2} \log  {(p-k)^2 + \mu^2 \over (p+k)^2 + \mu^2}\,
 \right] \ .
\eea
The mass gap equation (\ref{fixedpointeq})
for the running mass $m(p)$ is a non-linear
integral equation with a nasty cancellation
of Infrared divergences 
\cite{Adler:1984ri,Bicudo:2003cy,LlanesEstrada:1999uh}.
We devise a new method
with a rational ansatz,
and with relaxation
\cite{Bicudo:2010qp},
to get a maximum
precision in the IR
where the equation extremely
large cancellations occur.
Since the current quark masses of the six standard flavours $u, \, d, \, s, \, c , \, b, \, t$ span over five orders of magnitude from 1.5 MeV to 171 GeV, we develop an accurate numerical method to study the running quark mass gap and the quark vacuum energy density from very small to very large current quark masses. 
The solution $m(p)$  is shown in Fig. \ref{fitmassgap} for a vanishing momentum
$p=0$.

At finite $T$, one only has to change the string tension to the finite T
string tension $\sigma(T)$
of Eq. (\ref{eqformagnetization}), for different quark masses
\cite{Bicudo:2010hg}, 
and also to replace an integral in $p^0$ 
by a discrete sum in Matsubara Frequencies. Both are equivalent to a reduction in the
string tension, $\sigma \to \sigma^*$ and thus all we have to do is to solve the mass gap
equation in units of $\sigma^*$ .
The results are depicted in Fig. \ref{fitmassgap}.
Thus at vanishing $m_0$ we have a chiral symmetry phase transition,
and at finite $m_0$ we have a crossover,
that gets weaker and weaker when $m_0$ increases. This is also sketched
in Fig. \ref{fitmassgap}.

\section{Infrared regularization of the linear confining potential and
the Matsubara sum}

Notice that in the case of a linear potential, divergent in the
infrared, the Fourier transform needs an infrared
regulator $\mu$ eventually vanishing. We illustrate two
possible infrared regularizations of the linear potential
if Fig. \ref{linearregu}.

A possible regularization of the linear potential is,
\be
V(r)= - \sigma {e^{- \mu \, r} \over \mu} \simeq - {\sigma \over \mu} + \sigma r \ ,
\label{IRdivpot}
\ee
corresponding to a model of confinement where the quark-antiquark system has
an infinite binding energy ${\sigma \over \mu}$at the origin $r=0$, is monotonous 
and only vanishes at an arbitrarily large distance. 
This potential has a simple three-dimensional Fourier transform,
\bea
V(k) &=& \int_0^\infty  dr  \, { 4 \pi r \sin( k r ) \over k }  \, V( r)  
\nonumber \\
&=&  \sigma { - 8 \pi \over ( k^2 + \mu^2)^2   }
\eea
and this is the most common form of the linear potential in momentum space
utilized in the literature. Notice that this is infrared divergent due to the $-1/\mu$ infinite binding energy
in the limit where the regulator $\mu \to 0$.

\begin{figure}[t!]
\hspace{-2cm}
\begin{flushleft}
\includegraphics[width=1.1\columnwidth]{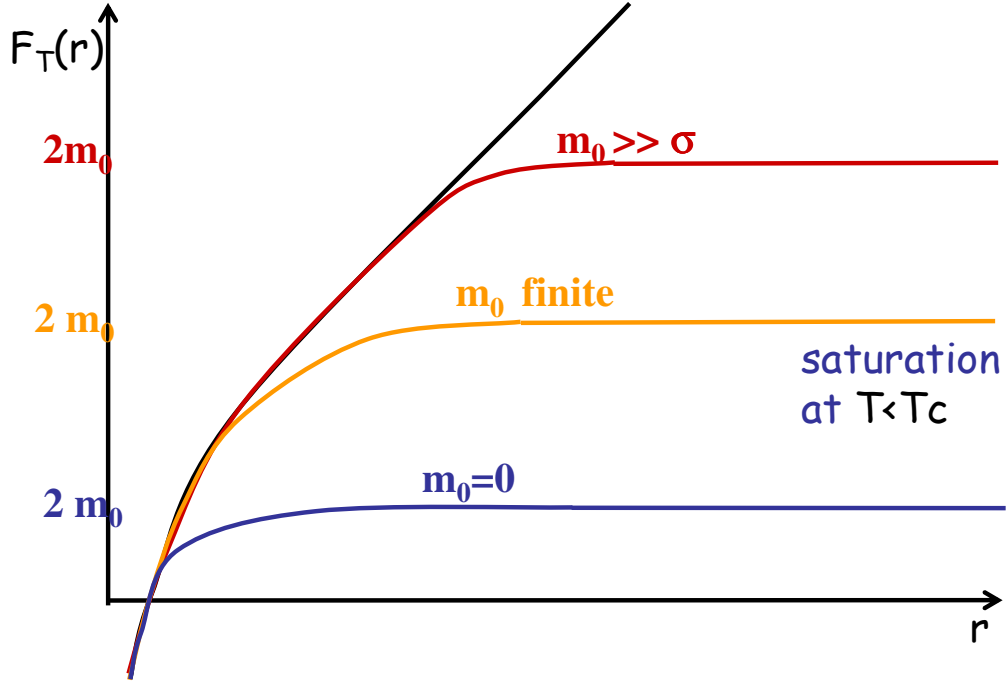}
\end{flushleft}
\vspace{-2cm}
\caption{Sketch of the saturation of confinement at the energy of
two heavy-light groundstate mesons. 
}
\label{sturationconfinement1}
\end{figure}

If we want to avoid the infinite binding energy we should use a different
regularization of the linear potential, also vanishing when $r \to \infty$ but
not monotonous since it grows linearly at the origin starting with $V(0)=0$,
\be
V(r)= \sigma r \, e^{- \mu \, r} 
\label{IRfinpot}
\ee
where the Fourier transform,
\be
V(k) =\sigma { - 8 \pi \over ( k^2 + \mu^2)^2   } + \sigma {32 \pi \mu^2 \over
 ( k^2 + \mu^2)^3  } 
\ee
is such that the integrals in $k$ no longer diverge. For instance,
\be
\int_{-\infty}^{+ \infty} k^2 dk V(k) = 0
\ee
since this is proportional to $V(0)=0$. The new term in the potential
$ {32 \pi \mu^2 \over
 ( k^2 + \mu^2)^3  } $ is equal to $ (2 \pi)^3 \delta^3(k) / \mu$ in
 the limit $\mu \to 0$, and this potential is infrared finite.
Both the potentials in Eqs. (\ref{IRdivpot}) and  (\ref{IRfinpot}) are
illustrated in Fig. \ref{linearregu}. 

In the vanishing temperature limit
$T=0$ the different regularizations lead to the same physical results since
any constant term in a density-density interaction has no effect in the quark 
running mass $m(p)$ or in the hadron spectrum
\cite{Bicudo:2010qp}.
The regularizations only contribute to the potential and the one-quark energy, 
but it occurs that these contributions exactly cancel in the chiral order parameters,
and in the hadron spectrum.

\begin{figure}[t!]
\center{
\includegraphics[width=1.1\columnwidth]{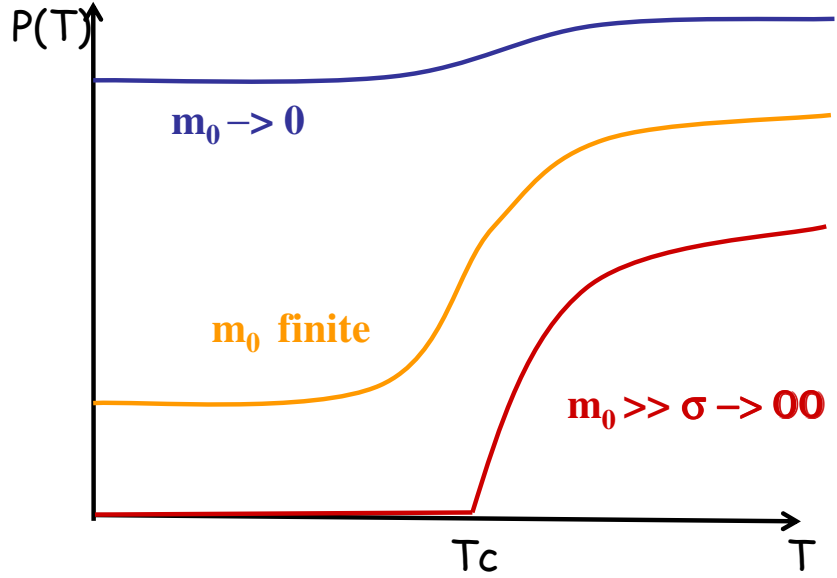}
}
\vspace{-2cm}
\caption{Sketch the order parameter $P$ polyakov loop for
the quark antiquark-system, as implied by the saturation of
the energy of the quark-antiquark system.
}
\label{sturationconfinement2}
\end{figure}

However at $T \neq 0$ the two different regularizations may lead
to different physical results.

In the $T=0$ mass gap equation or Schwinger Dyson equation, 
we have the Minkowski space integral in $p^0$ of the quark propagator pole of
Eq. (\ref{minkowski}) and this is 
equivalent to an integral in $p^4$ in Euclidian space after a Wick rotation in the 
Argand space,  
\be
\int_{- \infty}^{+ \infty} {i d \, p^4 \over 2 \pi}
{ i \over i p^4 - E(\mbf p) + i \epsilon}    = -{1 \over 2 } \ .
\label{euclidian}
\ee 
real axis the path corresponds to $z=p^0$ and
in the imaginary axis the path corresponds to $z = i \, p^4$.

Notice that the integral in Eq. (\ref{euclidian}) is only identical to the one in Eq. (\ref{minkowski})  if the one quark
energy $E(p) >0 $. If $E(p) <0 $ the integral of Eq. (\ref{euclidian}) changes sign, and this is consistent with the pole
moving from the fourth quadrant to the third quadrant of the Argand plane. While the integral in Eq. (\ref{minkowski}) 
is insensitive to this translation of the pole, the Wick rotation leads to a pole correction. For simplicity,
we choose to work with positive one quark energies only $E(p) >0 $.

In finite temperature $T$ and density $\rho$, the continuous euclidian space 
integration of eq. (\ref{euclidian})
is extended to the sum in Matsubara frequencies,
\be
\sum_{n=- \infty}^{+ \infty} i \, k T
{ i  \over i \, (2 n +1) \pi \, k T - \left[E(\mbf p) \right]} \ .
\ee
It is clear that in the vanishing temperature and density $kT  << E $ limit one gets back the initial 
Euclidean space integral of eq. (\ref{euclidian}), when the Matsubara sum approaches 
the continuum integration with $ k T \to {d P^4 \over 2 \pi}$.

\begin{figure}[t!]
\vspace{-4cm}
\begin{flushleft}
{\hspace{2cm} \includegraphics[width=1.2\columnwidth]{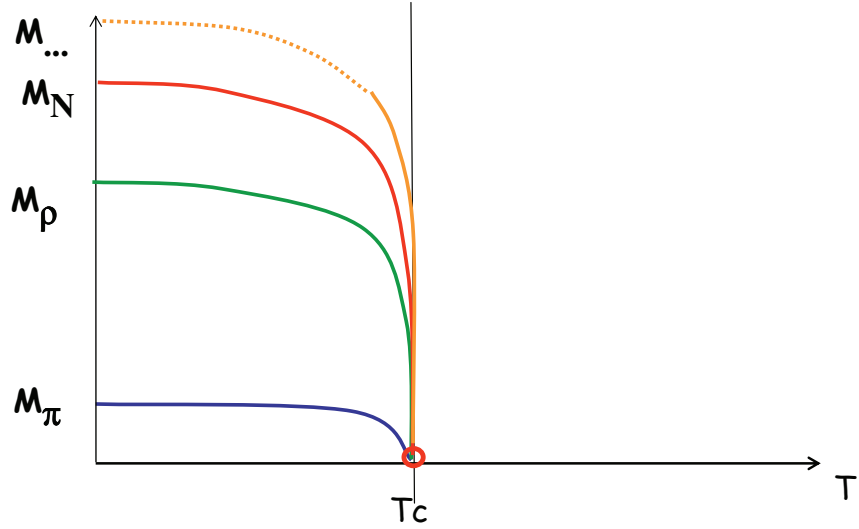}}
\end{flushleft}
\vspace{-2cm}
\caption{We sketch the light hadron spectrum. Notice that with $T \neq 0$, 
the light masses and wavefunctions do essentially scale with $\sigma(T)$, and thus also essentially vanish at $T=T_c$.
}
\label{lastfig2}
\end{figure}

Notice that at $T\neq 0$ the results depend on the constant in the potential.
Since the Matsubara sum depends on the one quark energy $E(p)$, and a 
constant shift of the potential added to the linear term, say  $V=\sigma(T) \, r-U_0$ 
$U_0>0$, affects the energy by a ${ 1 \over 2} \, U_0$,
thus weakening the finite $T$ effects scaling like $KT \over E$. On the other hand
if the shift is small enough to produce at some $p$ a negative energy $E(p)<0$,
at vanishing density $\mu =0$, then the Matsubara sum could break down.

Thus we study here only one scenario for the constant potential shift $-U_0$,
present anyway in the different possible infrared regularizations of the linear potential.
Our scenario consists in a maximum energy, with   $U_0 \to \infty$ as in the standard
regularization of the linear potential of Eq. (\ref{IRdivpot}). In that case 
$E(p) \to \infty$ and the Matsubara sum is simply constant, producing always
the $T=0$ result of $- 1 \over 2$. Our results are shown in Figs.  \ref{massgapofTlights} 
and \ref{massgapofT}.

We don't explore this second possible scenario here, leaving it for a future study, 
consisting in having the minimal $U_0$, closer to the regularization 
of the linear potential of Eq. (\ref{IRfinpot}), just sufficient  to cancel the one quark energy
at vanishing momentum $E(0)=0$. In that case the Matsubara sum makes a larger difference,
interpolating between $0$ at vanishing momentum and $- {1 \over2}$ at large
momentum.  But we expect that, whatever the contribution of the Matsubara sum is,
the temperature dependence of the string tension $\sigma(T)$ is the dominant finite $T$ 
effect, and thus for a first exploration our first scenario is sufficient.

\section{Chiral symmetry and confinement
crossovers with a finite current quark mass}

We now study whether the two main phase transitions in the QCD phase diagram, 
confinement and chiral symmetry breaking, have two different two critical points 
or a coincident one. Confinement drives chiral symmetry breaking, and at small 
density both transitions are a crossover, and not a first or second order phase 
transition due to the finite quark mass. 

However the quark mass affects differently these two phase transitions in the QCD. 
In what concerns confinement, the linear confining quark-antiquark potential
saturates when the string breaks at the threshold for the creation of 
a quark-antiquark pair.
Thus the free energy $F(0)$ of a single static quark is not infinite, 
it is the energy of the string saturation. The saturation energy is 
of the order of the mass of a meson i. e. of $ 2 m_0$.
For the Polyakov loop we get,
\be
P(0) \simeq N e^{ - 2 m_0 / T} \ .
\ee
Thus at infinite $m_0$ we have a confining phase transition,
while at finite $m_0$ we have a crossover,
that gets weaker and weaker when $m_0$ decreases.
This is sketched in Figs. \ref{sturationconfinement1} and \ref{sturationconfinement2}.

\begin{figure}[t!]
\vspace{-4cm}
\begin{center}
{\hspace{-2cm}\includegraphics[width=1.1\columnwidth]{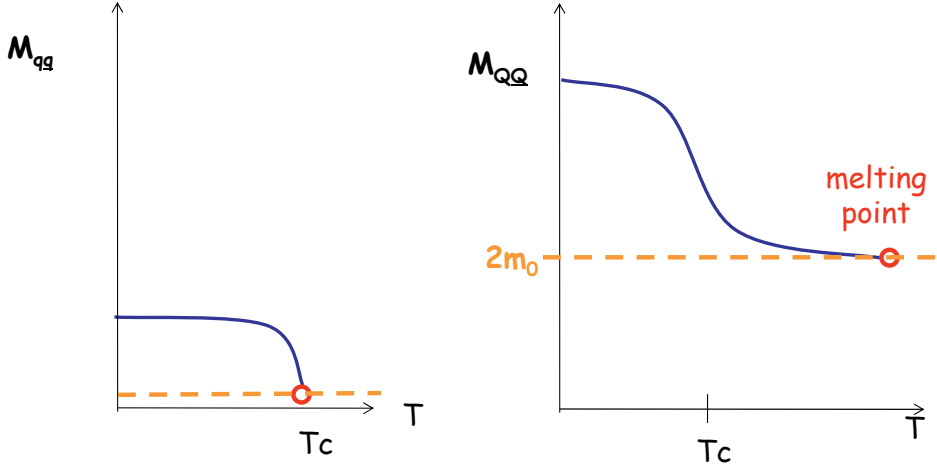}}
\end{center}
\vspace{-2cm}
\caption{
We sketch the behaviour of the meson spectra with temperature. The light hadron 
masses $M_{q \bar q}$ are dominated 
by the linear confining potential, and essentially vanish at $T_c$.
On the other hand, the heavy hadron masses 
$M_{Q \bar Q}$ 
are also significantly bound by the Coulomb potential,
providing enough binding up to a melting temperature above Tc.
}
\label{lastfig1}
\end{figure}

Since the finite current quark mass affects in opposite ways the crossover
for confinement and the one for chiral symmetry, we conjecture that at finite $T$ and $\mu$
there are not only one but two critical points (a point where a crossover
separates from a phase transition). 
Since for the light $u$ and $d$ quarks 
the current mass $m_0$ is small, we expect the crossover for chiral symmetry restoration 
critical to
be closer to the $\mu=0$ vertical axis, and the crossover for deconfinement
to go deeper into the finite $\mu$ region of the critical curve in the QCD
phase diagram depicted in Fig. \ref{CBM}.

We now compute the hadron spectrum, in particular the meson spectra. 
In what concerns the light meson masses $M_{q \bar q}$ 
are dominated by the linear confining potential.
At $T_c$, the string tension $\sigma(T)$ vanishes, 
the confining potential disappears, 
and thus all light hadrons decrease their mass until they melt at $T=T_c$.  
Notice that chiral symmetry is not entirely restored because 
$m_0/ \sqrt{\sigma(T)}$  increases with $T$. 
Nevertheless,  with $T \neq 0$, the masses and wavefunctions do essentially 
scale with $\sigma(T)$, and thus also essentially vanish at $T=T_c$. This is
sketched in Fig. \ref{lastfig2}. 
A comparable vanishing of light meson masses also occurs in the
sigma model
\cite{Elina}.

In what concerns the heavy meson masses $M_{Q \bar Q}$,
the Coulomb potential is also relevant, and thus the groundstates of the
heavy mesons still bind above Tc.
Thus while the light mesons melt at $T=T_c$, the heavy meson
groundstates melt at a $T > T_c$
\cite{Bicudo:2008gs,Bicudo:2009pb},
as depicted in Fig. \ref{lastfig1}.

\section{Outlook}

We remark that the pure gauge string tension $\sigma (T)$ is well fitted 
by the condensed matter physics magnetization curve $\chi$, 
and we utilize it.

We compute the dynamically generated quark mass $m(p)$ , 
solving the mass gap equation both for finite current quark masses $m_0$ 
and for finite $T$. The finite current quark masses turn both the confinement 
and the chiral symmetry phase transitions into two different crossovers.  

We qualitatively study the full spectra of light hadrons at finite $T$, 
including the excited spectra, and conclude that 
the light hadron masses essentially vanish at $T=T_c$.
The light hadrons all melt at $T=T_c$,
since the masses and wavefunctions essentially scale with $\sigma (T)$.

We soon plan to complete the part of this work which is only sketched here, {\em i. e.} to
compute the excited hadron spectrum at finite $T$, continuing the work initiated with
 Tim Van Cauteren, Marco Cardoso, Nuno Cardoso and Felipe Llanes-Estrada
 \cite{Bicudo:2009cr},
and to compute and compare
the crossover curves for the chiral symmetry restoration and for the deconfinement at
finite $T$.


\begin{thebibliography}{99}


\bibitem{CBM} 
CBM Progress Report, publicly available at
http://www.gsi.de/fair/experiments/CBM, (2009).

  
\bibitem{Aoki:2006we}
  Y.~Aoki, G.~Endrodi, Z.~Fodor, S.~D.~Katz and K.~K.~Szabo,
  Nature {\bf 443}, 675 (2006)
  [arXiv:hep-lat/0611014].

  
\bibitem{Bicudo:2009cr}
  P.~Bicudo, M.~Cardoso, T.~Van Cauteren and F.~J.~Llanes-Estrada,
  Phys.\ Rev.\ Lett.\  {\bf 103}, 092003 (2009)
  [arXiv:0902.3613 [hep-ph]].
  
\bibitem{Bicudo:2009hm}
  P.~Bicudo,
  Phys.\ Rev.\  D {\bf 81}, 014011 (2010)
  [arXiv:0904.0030 [hep-ph]].
  
  
\bibitem{Doring:2007uh}
  M.~Doring, K.~Hubner, O.~Kaczmarek and F.~Karsch,
  Phys.\ Rev.\  D {\bf 75}, 054504 (2007)
  [arXiv:hep-lat/0702009].

\bibitem{Hubner:2007qh}
  K.~Hubner, F.~Karsch, O.~Kaczmarek and O.~Vogt,
  arXiv:0710.5147 [hep-lat].
  
\bibitem{Kaczmarek:2005ui}
  O.~Kaczmarek and F.~Zantow,
  Phys.\ Rev.\  D {\bf 71}, 114510 (2005)
  [arXiv:hep-lat/0503017].
  
\bibitem{Kaczmarek:2005gi}
  O.~Kaczmarek and F.~Zantow,
  arXiv:hep-lat/0506019.

\bibitem{Kaczmarek:2005zp}
  O.~Kaczmarek and F.~Zantow,
  PoS {\bf LAT2005}, 192 (2006)
  [arXiv:hep-lat/0510094].
  

  
\bibitem{Kaczmarek:1999mm}
  O.~Kaczmarek, F.~Karsch, E.~Laermann and M.~Lutgemeier,
  Phys.\ Rev.\  D {\bf 62}, 034021 (2000)
  [arXiv:hep-lat/9908010].
  
\bibitem{FeynmanLS}
  R.~Feynamn, R.~Leighton, M.~Sands,
  "The Feynman Lectures on Physics", Vol II, chap. 36 "Ferromagnetism",
 published by Addison Wesley Publishing Company, Reading, Massachussets, ISBN 0-201-02117-x (1964).
 
 
 
\bibitem{Bicudo:2008kc}
  P.~Bicudo,
  Phys.\ Rev.\  D {\bf 79}, 094030 (2009)
  [arXiv:0811.0407 [hep-ph]].

\bibitem{TDLee}
T.D. Lee, Particle Physics and Introduction to Field Theory, (Harwood Academic Pub-
lishers, New York, 1981).


\bibitem{Szczepaniak:1995cw}
  A.~Szczepaniak, E.~S.~Swanson, C.~R.~Ji and S.~R.~Cotanch,
  Phys.\ Rev.\ Lett.\  {\bf 76}, 2011 (1996)
  [arXiv:hep-ph/9511422].

\bibitem{Szczepaniak:1996gb}
  A.~P.~Szczepaniak and E.~S.~Swanson,
  Phys.\ Rev.\  D {\bf 55}, 1578 (1997)
  [arXiv:hep-ph/9609525].



 

\bibitem{Balitsky:1985iw}
  I.~I.~Balitsky,
  Nucl.\ Phys.\  B {\bf 254}, 166 (1985).
  
\bibitem{Dosch:1987sk}
  H.~G.~Dosch,
  Phys.\ Lett.\  B {\bf 190}, 177 (1987).
  
\bibitem{Dosch:1988ha}
  H.~G.~Dosch and Yu.~A.~Simonov,
  Phys.\ Lett.\  B {\bf 205}, 339 (1988).
  
  

\bibitem{Bicudo:1998bz}
  P.~Bicudo, N.~Brambilla, E.~Ribeiro and A.~Vairo,
  Phys.\ Lett.\  B {\bf 442}, 349 (1998)
  [arXiv:hep-ph/9807460].
  











\bibitem{Orsay1} A. Le Yaouanc, L. Oliver, O. Pene, J. C. Raynal, Phys. Lett.
{\bf 134B}, 249 (1984).

\bibitem{Orsay2} A. Amer, A. Le Yaouanc, L. Oliver, O. Pene and J.-C. Raynal, Phys. Rev. Lett.
{\bf 50}, 87 (1983).

\bibitem{Orsay3} A. Le Yaouanc, L. Oliver, O. Pene and J.-C. Raynal,
Phys. Rev. D {\bf 29}, 1233 (1984); 

\bibitem{Orsay4}
A.~Le Yaouanc, L.~Oliver, S.~Ono, O.~P\`ene  and J.~C.~Raynal,
  Phys.\ Rev.\ D {\bf 31}, 137 (1985).

\bibitem{Kalinowski}
Y. L. Kalinovsky, L. Kaschluhn and V. N. Pervushin, Phys. Lett. B {\bf 231}, 288 (1989).

\bibitem{Lisbon1}  
 P.~Bicudo, J.~E.~Ribeiro, Phys.\ Rev.\ D {\bf 42}, 1611 (1990);
Phys.\ Rev.\ D {\bf 42}, 1625 (1990); Phys.\ Rev.\ D {\bf 42}, 1635 (1990).




\bibitem{linear1} S. L. Adler, A. C. Davis, Nucl. Phys. B {\bf 244}, 469 (1984),


\bibitem{linear2}
P.~Bicudo, J.~E.~Ribeiro and J.~Rodrigues, Phys.\ Rev.\ C {\bf 52}, 2144 (1995).

\bibitem{linear3}
R. Horvat, D. Kekez, D. Palle and D. Klabucar, Z. Phys. C {\bf 68}, 303 (1995).


\bibitem{linear4}
F. J. Llanes-Estrada, S. R. Cotanch, Phys. Rev. Lett.  {\bf 84}, 1102 (2000).

\bibitem{Wagenbrunn:2007ie}
  R.~F.~Wagenbrunn and L.~Y.~Glozman,
  Phys.\ Rev.\  D {\bf 75}, 036007 (2007)
  [arXiv:hep-ph/0701039].



  
\bibitem{Bicudo:2010qp}
  P.~Bicudo,
  arXiv:1007.2044 [hep-ph].
 
\bibitem{Adler:1984ri}
  S.~L.~Adler and A.~C.~Davis,
  Nucl.\ Phys.\  B {\bf 244}, 469 (1984).
  
\bibitem{Bicudo:2003cy}
  P.~J.~A.~Bicudo and A.~V.~Nefediev,
  Phys.\ Rev.\  D {\bf 68}, 065021 (2003)
  [arXiv:hep-ph/0307302].
  
\bibitem{LlanesEstrada:1999uh}
  F.~J.~Llanes-Estrada and S.~R.~Cotanch,
  Phys.\ Rev.\ Lett.\  {\bf 84}, 1102 (2000)
  [arXiv:hep-ph/9906359].



\bibitem{Bicudo:2010hg}
  P.~Bicudo,
  Phys.\ Rev.\  {\bf D82}, 034507 (2010).
  [arXiv:1003.0936 [hep-lat]].
  
  
\bibitem{Elina}
E. Seel, S. Strüber, F. Giacosa,  D. H. Rischke
"Chiral symmetry restoration in linear and 
nonlinear O(N) models  within the auxiliary
field method,"
invited talk at Excited QCD 2011,
Les Houches, France, February 2011.
  
\bibitem{Bicudo:2008gs}
  P.~Bicudo, M.~Cardoso, P.~Santos, J.~Seixas,
    [arXiv:0804.4225 [hep-ph]].
  
\bibitem{Bicudo:2009pb}
  P.~Bicudo, J.~Seixas, M.~Cardoso,
    [arXiv:0906.2676 [hep-ph]].
  
  
  
  
\end{thebibliography}
\end{document}